\documentclass{epl}
\usepackage{graphicx}% Include figure files
\usepackage{dcolumn}% Align table columns on decimal point
\usepackage{bm}% bold math
\usepackage{amssymb}

\begin{document}

\title{Hanle effect in transport through quantum dots coupled to ferromagnetic leads}
\shorttitle{Hanle effect in quantum dots}

\author{M. Braun\inst{1} \and J. K\"onig\inst{1} \and J. Martinek\inst{2,3,4}}
\institute{
  \inst{1} Institut f\"ur Theoretische Physik III - Ruhr-Universit\"at Bochum, 44780 Bochum, Germany\\
  \inst{2} Institut f\"ur Theoretische Festk\"orperphysik - Universit\"at Karlsruhe, 76128 Karlsruhe, Germany\\
  \inst{3} Institute of Molecular Physics - Polish Academy of Science, 60-179 Pozna\'n, Poland\\
  \inst{4} Institute for Materials Research, Tohoku University, Sendai 980-8577, Japan}

\date{\today}

%\pacs{72.25.Mk}{Spin transport through interfaces}

\maketitle

\begin{abstract}
We suggest a series of transport experiments on spin precession in
quantum dots coupled to one or two ferromagnetic leads.
Dot spin states are created by spin injection and analyzed via the linear 
conductance through the dot, while an applied magnetic field gives rise to 
the Hanle effect.
Such a Hanle experiment can be used to determine the spin lifetime 
in the quantum dot, to measure the spin injection efficiency into the 
dot, as well as proving the existence of intrinsic spin 
precession which is driven by the Coulomb interaction.
\end{abstract}

Recent progress in nanofabrication technology opens the possibility
for spintronic devices based on coherent manipulation of single spins
in quantum dots \cite{quantum_computation}.
Thereby the spatial confinement of the dot electrons suppresses spin
decoherence \cite{SO}.
While longitudinal spin relaxation times $T_1$ up to microseconds were
measured \cite{fujisawa,kouwenhoven}, the size of the spin decoherence
time $T_2$ is still an open question.

The spin coherence time $T_2$ can be accessed in different experiments, 
for example by ESR techniques \cite{ESR}, or by the Hanle effect, {\it i.e.},
the decrease of spin accumulation in the quantum dot due to precession 
about a static magnetic field.
The optical realization of such a Hanle experiment involves the measurement of
the fluorescent emission of polarized light from semiconductor quantum dots \cite{hanle_in_dots}. 
But this method requires an ensemble of spins, so the 
total signal varies with the spin dephasing time $T_2^\star$ rather than the 
decoherence time $T_2 > T_2^\star$.  

To void this ensemble averaging, we suggest to measure the Hanle effect 
in transport through an individual single level quantum dot. 
For preparation of the initial spin state, electronic spin injection 
from ferromagnetic leads into the dot can be used. This has been demonstrated
in metallic bulk systems \cite{johnson}, quantum dots \cite{spin-injection},
Zener diodes \cite{hanle_in_Zener_diode}, and metallic grains \cite{wees}.
For the detection of the accumulated spin, we propose magnetoresistance 
measurements of the device as already shown for metallic systems \cite{wees,johnson}. 
So an all electrical experimental setup can be used to observe the 
reduction of spin accumulation in a single quantum dot by an external magnetic field, 
which is a direct measure of the product of precession frequency $\omega$ and spin lifetime.

In a recent experiment Zhang {\em et. al.} \cite{grain_experiment} realized 
this kind of setup but with a whole layer of aluminum dots in a tunnel 
junction between two Co electrodes with many levels participating in transport 
in each dot rather than an individual quantum dot with only a single level
contributing to the current.
Even so the measurements involve averaging over different 
realizations of the dots, multi levels and local magnetizations, they clearly 
observe a Hanle resonance in the magnetoresistance of the device.

\begin{figure}[h!]
\begin{center}
\includegraphics[angle=0,width=0.8\columnwidth]{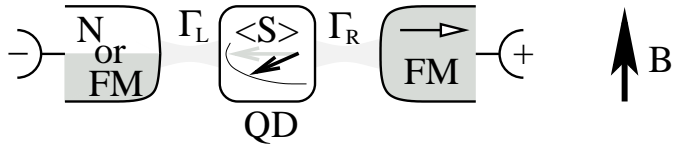}
\caption{\label{sketch}
Quantum dot connected to one or two ferromagnetic leads. Spin dependent tunnel rates lead to spin accumulation on the dot. This accumulated spin precesses in the external applied magnetic field.}
  \end{center}
\end{figure}

For our theoretical model we consider only one single-level quantum 
dot with level energy $\varepsilon$, measured relative to the Fermi energy of the leads and
tunable by a gate electrode as sketched in fig.\ref{sketch}.
The dot level can be empty, singly occupied, or doubly occupied.
The Coulomb interaction on the dot is accounted for by the charging energy $U$
for double occupancy.
The dot is coupled to source and drain electrodes by tunnel contacts.
The coupling strength to the left and right lead is characterized by the
intrinsic line width $\Gamma_r$ with $r = {\rm L}, {\rm R}$.
We study sequential-tunneling regime, which yields
$k_{\rm B}T \gg \Gamma_r$.
The electrodes may be nonmagnetic or ferromagnetic, described by their
degree of spin polarization $p_r$ with $0\le p_r \le 1$ and magnetization
direction $\hat{\bm n}_r$.
Further possible realizations of such systems include molecular \cite{ralph} or
carbon nanotube \cite{CNT} junctions, grains in
nano-constrictions \cite{large_fields} and junctions \cite{grain_experiment}, or surface impurities contacted by an STM tip \cite{STM}.

A full theoretical description of the spin dynamics and its implication on
sequential-tunneling transport has been derived in ref.~\cite{master}
within a diagrammatic transport theory.
Here, we use this theory to analyze the Hanle effect.
To keep the discussion transparent, we assume symmetric coupling constants,
$\Gamma_{\rm L}=\Gamma_{\rm R}=\Gamma/2$ and consider only the regime of
linear transport, {\em i.e.} bias voltages smaller than temperature $eV\ll k_{\rm B}T$.
The spin accumulation on the quantum dot is described by the quantum 
statistical value $\bm S$, which has the co-domain $0 < |\bm S| < \hbar/2$.
The dynamics of the dot spin $\bm S$ is governed by a Bloch-like equation
\begin{equation}
\label{S}
   \frac{d\bm S}{dt}= \sum_{r=\rm L/R}\left( \frac{\hbar}{2e} I_r p_r \hat{\bm n}_r
       -\frac{\bm S-p^2(\hat{\bm n}_r\cdot \bm S)\hat{\bm n}_r}{\tau_{c,r}}\right)
       +{\bm S} \times {\bm \omega}-\frac{\bm S}{\tau_{\rm rel}}
       \, .
\end{equation}
The first term describes spin accumulation due to spin-polarized currents from
or into ferromagnetic leads. In the steady state, $I_{\rm L}=-I_{\rm R}$, so 
the spin does accumulate in the direction 
$p_{\rm L} \hat{\bm n}_{\rm L}-p_{\rm R} \hat{\bm n}_{\rm R}$. 

The second term describes the relaxation of the dot spin due to coupling 
to the leads.
Since neither an empty nor a doubly-occupied dot can bear a net 
spin, the relaxation time is exactly the life time of the 
single-occupation dot state.
The time scale for tunneling of an electron to or from the electrode $r$ 
is given by $\tau_{c,r}^{-1}= \Gamma_r/\hbar(1-f_r(\varepsilon)+f_r(\varepsilon+U))$, 
where $f_r(\varepsilon)$ is the Fermi function of the lead. 
The total life time of the single-occupation state for a weak bias voltage is given by 
$\tau_{\rm c}^{-1}=\sum_r \tau_{c,r}^{-1}= \Gamma/\hbar(1-f(\varepsilon)+f(\varepsilon+U))$.
Together with the phenomenological spin-relaxation rate 
$\tau_{\rm{rel}}^{-1}$, describing decoherence {\it e.g.}, due to spin-orbit 
coupling or hyperfine interaction within the quantum dot, the total spin 
coherence time of the dot spin is 
\begin{eqnarray}\label{tau}
  \left(\tau_{\rm{s}}\right)^{-1} = \left(\tau_{\rm{rel}}\right)^{-1}
  + \left(\tau_{\rm{c}}\right)^{-1} \, .
\end{eqnarray}
Quantum information processing with quantum-dot spins is anticipated to be
operated in the Coulomb-blockade regime, where $\tau_{\rm{c}}$ is large and
$T_2$ is limited by $\tau_{\rm{rel}}$.
Below we will propose a scheme to measure $\tau_{\rm{s}}$ in the opposite,
the sequential-tunneling regime.
Therefore, in order to estimate the relevant $T_2$ for quantum-computing
applications from the measured $\tau_{\rm{s}}$, one has to subtract the
influence of $\tau_{\rm{c}}$.

The third term in eq.~(\ref{S}) describes precession of the quantum-dot spin
about an effective magnetic field.
This includes an externally-applied magnetic field ${\bm B}$ and the exchange 
interaction to the ferromagnetic leads.
In the presence of strong Coulomb interaction on the dot, the tunnel coupling 
to ferromagnetic electrodes renormalizes the energy 
levels spin dependent \cite{master,Kondo}. This renormalization leads also to 
spin precession which can be described by a magentic-like exchange field. 
In total, we get
\begin{equation} \label{omega}
 {\bm \omega} = {\bm\omega}_{\rm{B}}
+{\bm\omega}_{\rm{x}}\, ,
\end{equation}
with $\hbar {\bm\omega}_{\rm{B}} = g\mu_{\rm{B}}{\bm B}$
and the exchange contribution
${\bm\omega}_{\rm{x}}=A  \Gamma [p_{\rm{L}} \hat {\bm n}_{\rm{L}}
    + p_{\rm{R}} \hat {\bm n}_{\rm{R}}]$.
In the linear-response regime, the $\varepsilon$ dependent 
numerical factor $A$ is determined by the integral
$A =(1/2\pi)\int' dE[\left(E-\varepsilon-U\right)^{-1}-
\left(E-\varepsilon\right)^{-1}] f(E)$.

Since both, the accumulation and the relaxation 
term in eq.(\ref{S})  transfer spin through the 
tunnel barrier, the partition in accumulation and 
damping is to some extend arbitrary. The current 
choice has the advantage, that the accumulation 
term is direct proportional to the electrical current.
On the other hand the interpretation in ref. \cite{master} 
gives a more intuitive isotropic damping term proportional 
to $\bm S$. 

In the following calculation we consider the parameter regime 
$\Gamma \approx \hbar \omega_B \ll eV\ll k_BT$. 
The magnetic field $\hbar \omega_B$ must be comparable
to $\Gamma$, to observe coherent rotation of the dot
spin rather than incoherent transport through 
Zeeman-splitted levels. Furthermore, if 
$\hbar {\bm \omega}_{\rm B}\lesssim \Gamma$, we can 
neglect the influence of the Zeeman splitting on the 
transition rates, since these corrections would be 
of the order $\Gamma B\approx\Gamma^2$, comparable to
cotunneling. The condition $\hbar \omega_B \ll eV$ 
ensures that the electrical spin injection dominates 
the spin dynamics rather than the equilibrium spin 
due to the Zeeman splitting. 
We emphasize that, although the latter condition requires a finite
transport voltage, we can calculate the conductance in linear response
as long as $eV\ll k_{\rm B}T$.

\begin{figure}[h!]
\begin{center}
\includegraphics[angle=-90,width=0.8\columnwidth]{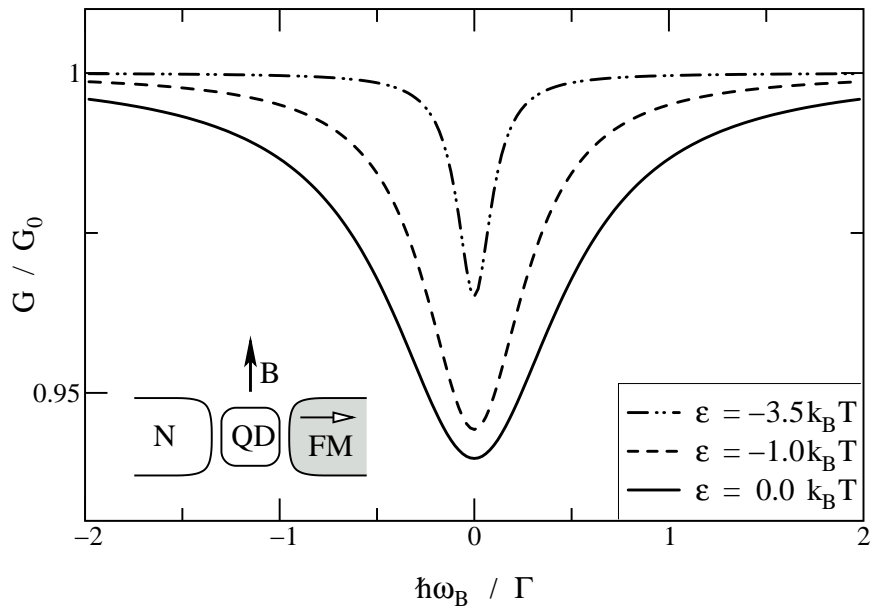}
\caption{\label{onelead}
  The differential conductance as function of the magnetic field applied
  transverse to the ferromagnetic lead fior different gate voltages. 
  By shifting the dot into Coulomb blockade, the electron dwell time of the 
  electrons increase. Since the electrons have then more time to precess and to relax inside the dot, 
 width of the Hanle resonance as well as the depth decreases.
  The parameters are $p=0.5$, $U=7k_{\rm B}T$,
  $\tau_{\rm rel}=20\hbar/\Gamma$, and level positions 
  $\varepsilon/k_{\rm B}T=-3.5$ (dot-dashed), $-1.5$ (dashed), and $0$ 
  (solid).}
\end{center}
\end{figure}

In the rest of the paper we apply our formalism to three
different setups. The first one consists of a quantum
dot attached to one nonmagnetic and one ferromagnetic
electrode (fig.~\ref{onelead}), a geometry already
realized experimentally \cite{large_fields}.

When a transport voltage is applied, the quantum dot
becomes spin polarized. The direction of the accumulated
spin depends on the direction of current flow (the linear
conductance, though, does not).
In case the electrons are flowing from the unmagnetized
source lead (left) to the ferromagetic drain (right),
the dot spin will accumulate anti-parallel to the magnetization
direction $\hat {\bm n}_{\rm{R}}$
since the tunnel barrier to the drain is more transparent
for electrons polarized along $\hat {\bm n}_{\rm{R}}$ than
for those along $- \hat {\bm n}_{\rm{R}}$.
A transverse magnetic field rotates the dot spin away from
this anti-parallel position. The rotated spin has an increased
component along $\hat {\bm n}_{\rm{R}}$, {\it i.e.}, the electron
can more easily tunnel into the ferromagnetic drain electrode.
As a consequence, the conductance increases, which
defines the Hanle effect in transport.
Based on the transport theory in ref.~\cite{master},
we derive the the linear conductance of this setup to be
\begin{eqnarray}
  \label{cond1}
  \frac{G}{G_0} = 2-\left[1-\frac{p^2}{4}\,
    \frac{\tau_{\rm{s}}}{\tau_{\rm{c}}}\,
    \frac{1+[\hat{\bm n}_{\rm{R}}\cdot
      ({\bm \omega}_{\rm B} + {\bm \omega}_{\rm x})\tau_{\rm s}]^2}
	 {1+({\bm \omega}_{\rm B}+{\bm \omega}_{\rm x})^2\tau_{\rm s}^2}\right]^{-1}\,,
\end{eqnarray}
where $G_0={e^2}\,{P_1}/{\tau_{\rm c}}{k_{\rm B}T}$ is
the asymptotic value of the conductance for large magnetic field,
$B \rightarrow \infty$, for which the spin accumulation is
completely destroyed. The latter is proportional to $P_1$,
the equilibrium probability to find the dot occupied by a
single electron, and $\tau_{\rm c}$ is given by
$\tau_{\rm c}^{-1}=(\Gamma/\hbar) \left[1-f(\varepsilon)+
f(\varepsilon+U)\right]$.
Note that the dependence of the expression in eq.~(\ref{cond1})
on the magnetic field differs from the optically-measured Hanle
signal \cite{hanle_in_Zener_diode}, as a consequence of the different way to
probe the spin.

Results are shown in fig.~\ref{onelead}, where ${\bm \omega}_{\rm B} =
g\mu_{\rm B}{\bm B}/\hbar$ characterizes an external magnetic field
applied perpendicular to the direction of lead magnetization, and ${\bm \omega}_{\rm x}$
the exchange field (aligned parallel), such that
${\bm \omega} = {\bm \omega}_{\rm B} + {\bm \omega}_{\rm  x}$.
For not too large values of the spin polarization $p$ in the ferromagnet,
$(p^2/4)\tau_{\rm  s}/\tau_{\rm  c}$ is small, and eq.~(\ref{cond1}) can be
expanded in the latter quantity.
The dip in the relative conductance at zero magnetic field is, thus,
approximatively $(p^2/4)\tau_{\rm  s}/\tau_{\rm  c}$, which provides an
estimate of the degree $p$ of spin polarization in the ferromagnet. 
To maximize $\tau_{\rm  s}/ \tau_{\rm  c} \le 1$ one can tune the quantum-dot
level on resonance.
The widths $\Delta \omega_{\rm  B}$ of the dip as a function of the applied 
field is determined by the condition $\Delta \omega_{\rm B} \tau_{\rm s} =
\sqrt{1 + 2(\omega_{\rm x} \tau_{\rm s})^2}$.
The inverse of the measured width, $1/\Delta \omega_{\rm B}$, does therefore 
only provide a lower limit for the spin lifetime $\tau_{\rm s}$, since the 
exchange field modifies the line width. 

External field components parallel to the magnetization directions influence 
the result in exactly the same way as the exchange field 
${\bm \omega}_{\rm x}$ does, leading to a broadening of the 
Hanle resonance. 
In the experiment by Zhang {\em et. al.} \cite{grain_experiment}, 
the authors propose that stray fields leading to this kind of broadening 
mechanism are the major source of error for measuring $T_2$.
To reduce the influence of the stray fields, the dot-lead coupling 
can be increased. Since $\tau_{\rm s}$ is inverse to $\Gamma$, 
the accumulated spin becomes less sensitive to the magnetic fields 
for stronger couplings, making small stray fields insignificant.
Typical transport experiments \cite{ralph,large_fields} show current
plateaus of the order 10pA to 10nA in the I-V characteristic, which
yields $\Gamma/\hbar$ to be of the order of $10^8{\rm s}^{-1}$ to
$10^{11}{\rm s}^{-1}$.
The typical magnetic field strength required for the 
observation of the precession is then of the order 100$\mu$T to 100mT, 
which can exceed the stray fields of the ferromagnetic leads in 
appropriate probe designs \cite{wees,grain_experiment}.
It is worth to mention, that since the exchange field is also a 
linear function of $\Gamma$, its influence, which is proportional to 
$\omega_{\rm x} \tau_{\rm s}$, does not depend on the coupling strength.

\begin{figure}[h!]
\begin{center}
\includegraphics[angle=-90,width=0.8\columnwidth]{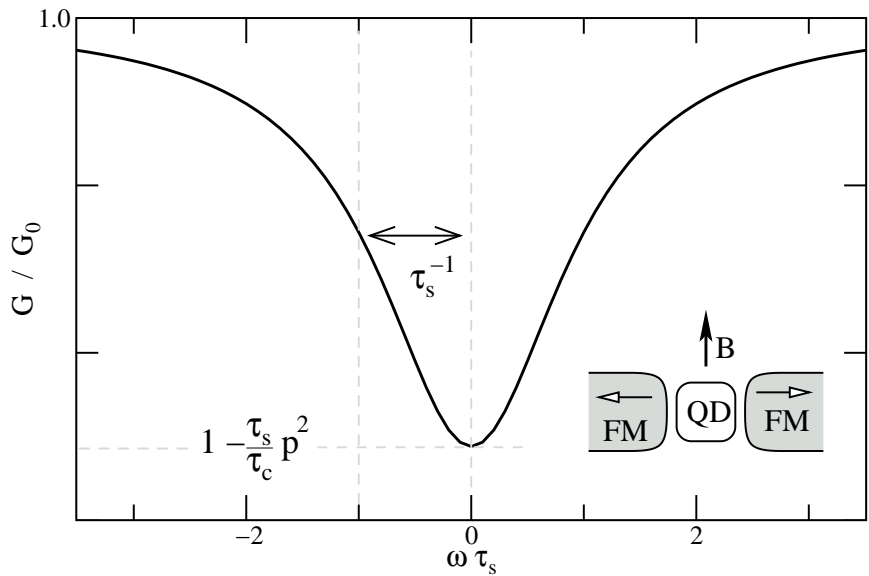}
\caption{\label{G_lowp}
  Differential conductance, for ferromagentic leads with anti-parallel
  magnetization, as a function of the magnetic field $\omega$ applied
  perpendicular to the accumulated spin. The half line width of the 
  Hanle resonance directly determines the spin coherence time $\tau_s$.}
\end{center}
\end{figure}

The structure discussed so far has the advantage that only one lead is
ferromagnetic, which might simplify the manufacturing procedure.
It is suitable to prove the existence of the spin precession analogous
to the optical Hanle effect.
The influence of the exchange field, though, makes it difficult to directly
determine $\tau_{\rm s}$.
We, therefore, turn now to a second setup which involves two ferromagentic
leads with magnetization directions anti-parallel to each other, 
see fig.\ref{G_lowp}.

For symmetric coupling $\Gamma_{\rm L}=\Gamma_{\rm R}$, equal degree of
polarization $p_{\rm L}=p_{\rm R}=p$ and in the linear-response regime, the
exchange field originating from the left and the right tunnel barrier cancel
out each other so $\omega_{\rm x}=0$, and the dot spin precesses only due to
the external magnetic field. The linear conductance, then, is
\begin{eqnarray}
  \frac{G}{G_0} &=& 1-p^2\frac{\tau_{\rm s}}{\tau_{\rm c}} \,
  \frac{1+(\frac{\hat{\bm n}_{\rm L}-\hat{\bm n}_{\rm R}}{2}{\bm \omega_{\rm B}}\tau_{\rm s})^2 } {1+({\bm\omega}_{\rm B} \tau_{\rm s})^2} \, .
\end{eqnarray}

If we assume the field to be aligned perpendicular to the lead
magnetizations (see fig.~\ref{G_lowp}), we find the Lorentzian
dependence on the external magnetic field that familiar from the
optical Hanle effect.
The depth of the dip is given by $p^2\tau_{\rm s}/\tau_{\rm c}$ while
the width of the dip in fig.~\ref{G_lowp} provides a direct
access to the spin lifetime $\tau_{\rm s}$.
Of course, the conversion of applied magnetic field to frequency 
requires the knowledge of the Lande factor $g$, which must be 
determined separately like in ref.~\cite{large_fields}.

\begin{figure}[h!]
\begin{center}
\includegraphics[angle=-90,width=1.0\columnwidth]{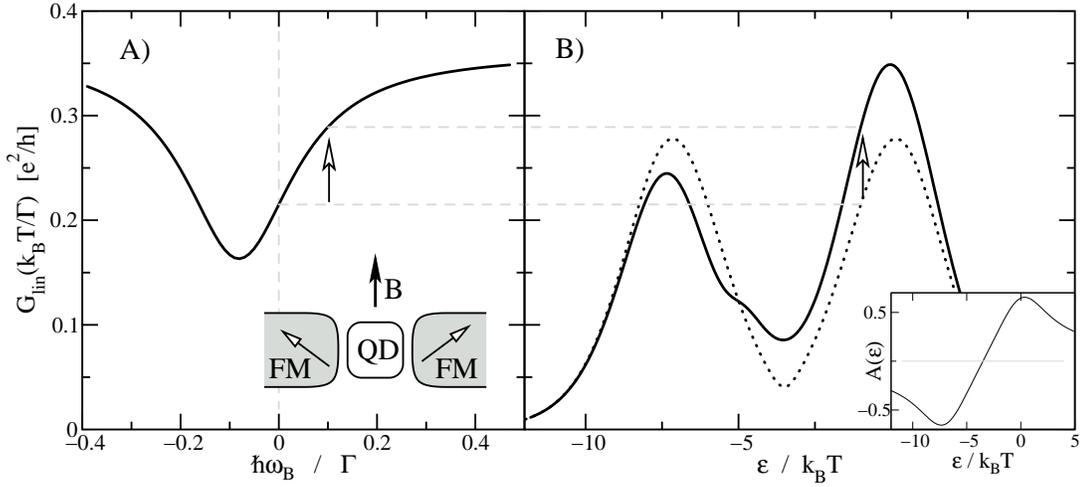}
\caption{\label{G_eps}
  Linear conductance of the dot for an applied external magnetic 
  field ${\bm B}$ along $\hat {\bm n}_{\rm L}+\hat {\bm n}_{\rm R}$.
  A) Linear conductance as a function of the applied field for
  $\varepsilon=0$.
  B) Linear conductance as a function of the level position
  $\varepsilon$ without external field (dashed-dot-dot) and for
  the applied field $\omega_{\rm B}=0.1\Gamma/\hbar$ (solid). 
  Further parameters are $\phi=3\pi/4$, $p=0.8$, $U=7k_{\rm B}T$, and
  $\tau_{\rm rel}=0$. The 
  vertical lines relate the Conductance increase of the dot at $\varepsilon=0$ for a magnetic field $\hbar\omega_B=0.1\Gamma$.}
  \end{center}
\end{figure}

Finally, we discuss the case of a non-collinear configuration
of the leads' magnetizations with a magnetic field applied along the direction
$\hat {\bm n}_{\rm L}+\hat {\bm n}_{\rm R}$ as shown in fig.~\ref{G_eps}.

In this case, both the exchange field and the external magnetic field are
pointing along $\hat {\bm n}_{\rm L}+\hat {\bm n}_{\rm R}$, perpendicular
to the accumulated spin.
The linear conductance is, then,
\begin{eqnarray}
  \label{GB}
 \frac{G}{ G_0}  &=&
  1-p^2\frac{\tau_{\rm s}}{\tau_{\rm c}} \,\,
  \frac{\sin^2\frac{\phi}{2}}{1+({\bm\omega}_{\rm B}+{\bm\omega}_{\rm x})^2 \tau_{\rm s}^2}\, ,\qquad
\end{eqnarray}
where $\phi=\sphericalangle(\hat{\bf n}_{\rm L};\hat{\bf n}_{\rm R})$ is the angle
enclosed by the leads' magnetization directions.

This setup allows for a stringent experimental verification of spin
precession due to the exchange field.
The conductance reaches its minimal value when the external magnetic field
has opposite direction and equal magnitude as the exchange field.
The shift of the minimum's position relative to $B=0$, thus, measures the 
exchange field, see fig.~\ref{G_eps}A. One can clearly separate the exchange 
field from possible stray fields of the leads by varying the gate voltage of 
the quantum dot. While the stray fields does not depend on the gate voltage, 
the exchange interaction does as plotted in the inset of fig.~\ref{G_eps}B.

In the flat band limit, the exchange field even changes its sign as a 
function of gate voltage, so by plotting the conductance as function 
of the gate voltage in Rig.~\ref{G_eps}B, we observe an increased 
conductance for one resonance peak, and a decreased for the 
other one.
At the intersection point of the conduction curves with and without 
external field, the absolute value of $\bm \omega$
remains unchanged but has opposite sign, {\it i.e.}, the exchange field at 
this point is $-g\mu_{\rm B}B/2$.

In summary, we suggest to measure the Hanle effect in transport through
quantum dots coupled to one or two ferromagnetic leads.
We propose schemes how to observe non-equilibrium spin accumulation, 
determine the dot-spin lifetime and verify
the existence of an intrinsic spin precession caused 
by Coulomb interaction.

We thank J. Barnas, B. Kubala, S. Maekawa, G. Sch\"on, and D. Urban for 
discussions.
This work was supported by the DFG under CFN, SFB 491, and GRK 726, the 
EC RTN on 'Spintronics', Project PBZ/KBN/044/P03/2001 and the EC Contract
G5MACT-2002-04049.

\end{document}